\documentclass[twocolumn]{revtex4-2}
\usepackage{graphicx}
\usepackage{dcolumn}
\usepackage{bm}
\usepackage{lipsum} 
\usepackage{xcolor} 
\usepackage{amsmath}
\usepackage{hyperref}
\usepackage[mathlines]{lineno}%

\newcommand{\removed}[1]{{}}

\newcommand{\dd}{\mathrm{d}}
\begin{document}

\title{Moving backward to go faster: Diatom-inspired sliding reveals  efficient modes of locomotion}

\author{Julien Le Dreff}
 \email{julien.le-dreff@polytechnique.edu}
\author{Blaise Delmotte}%
 \email{blaise.delmotte@cnrs.fr}
\affiliation{%
LadHyX, CNRS, Ecole Polytechnique, Institut Polytechnique de Paris, 91120 Palaiseau, France
}%

\begin{abstract}
Across biological scales, from sperm cells to whales, locomotion commonly relies on undulatory gaits, in which traveling deformation waves interact with the surrounding fluid to generate thrust opposite to the direction of wave propagation. In viscous environments, microorganism locomotion is classically understood in terms of undulatory bending of slender filaments such as flagella, with optimal propulsion achieved when the deformation wavelength is comparable to the swimmer length. Inspired by diatom colonies, we identify a fundamentally different swimming mechanism based on sliding between neighboring elements within a chain. We show that sliding between stacked elongated cells generates internal shear that drives propulsion opposite to classical undulatory swimming, while achieving higher speeds and greater energetic efficiency. Remarkably, optimal performance occurs at wavelengths much larger than the chain length and at cell aspect ratios consistent with those observed in natural diatom colonies, suggesting that hydrodynamic efficiency may constitute an evolutionary selective pressure in diatom chains. Together, these results identify sliding as a previously overlooked mode of locomotion in multicellular assemblies and suggest new design principles for efficient bio-inspired microswimmers and swarm robotic systems.\\

\textbf{Significance statement:} Many organisms swim by passing waves along their bodies or appendages, pushing against the surrounding fluid to move forward. At microscopic scales, this principle underlies flagellar swimming, which is most efficient when the wave pattern matches the swimmer’s size. We show that chains of elongated cells, inspired by diatom colonies, can move using a fundamentally different strategy based on sliding between neighboring cells. This sliding motion creates internal shear that propels the chain in a direction opposite to classical undulatory swimming, while achieving higher speeds and improved energetic efficiency. Optimal motion occurs at wavelengths much longer than the chain itself and for cell shapes consistent with those found in nature. These findings reveal sliding as an unexpected mode of locomotion in multicellular systems and suggest new design principles for bio-inspired microswimmers and collective robotic systems.
\end{abstract}

\maketitle


A vast number of species across nearly all kingdoms are capable of swimming, employing diverse strategies that have evolved to meet physical, ecological, and anatomical constraints. From the sinuous swimming of sperm cells to undulations of whales, undulatory gaits represent a fundamental and recurrent principle of aquatic locomotion across biological scales \cite{Brennen_review_1977,Gazzola_scaling_2014}. Despite vast differences in morphology and propulsion mechanisms, these organisms all rely on the same essential concept: a traveling deformation wave interacts with the surrounding fluid to generate thrust in the opposite direction \cite{Childress_1981}. At the microscale, swimming arises from the interplay between body deformation and viscous resistance. In this regime, characterized by low Reynolds number hydrodynamics, fluid inertia can be neglected, and propulsion can only emerge from nonreciprocal shape changes that break time-reversal symmetry \cite{Purcell_life_1977}. Flagellated cells such as spermatozoa achieve this through traveling bending waves that run along their slender tails, converting periodic oscillations into  forward motion \cite{gray1955propulsion, Brokaw_non-sinusoidal_1965, higdon1979hydrodynamic,dresdner1981relationships,  tam2011optimal}.

While much research on microscale swimming has focused on flagellated or ciliated cells, other microorganisms, such as diatoms, employ distinct mechanisms to achieve motion \cite{Brennen_review_1977,lauga_hydrodynamics_2009}. Diatoms are unicellular photosynthetic organisms that play a key role in both marine and freshwater ecosystems \cite{Armbrust_life_2009}. They are characterized by silica cell walls, known as frustules.
In many \removed{pennate} diatoms, the frustule features opposite longitudinal slits called raphes \cite{round1990diatoms}.
Despite the absence of any visible appendages, certain species of diatoms can move along surfaces via a mechanism known as gliding motility. 
This movement is thought to be driven by forces generated by an actin-myosin complex that interacts with an adhesive extracellular matrix secreted through the raphes \cite{poulsen1999diatom,BondocNaumovitz_functional_2025,Zhang_ice_2025,Yamaoka_colonial_2016} 



\begin{figure*}
\centering
\includegraphics[width=0.9\linewidth]{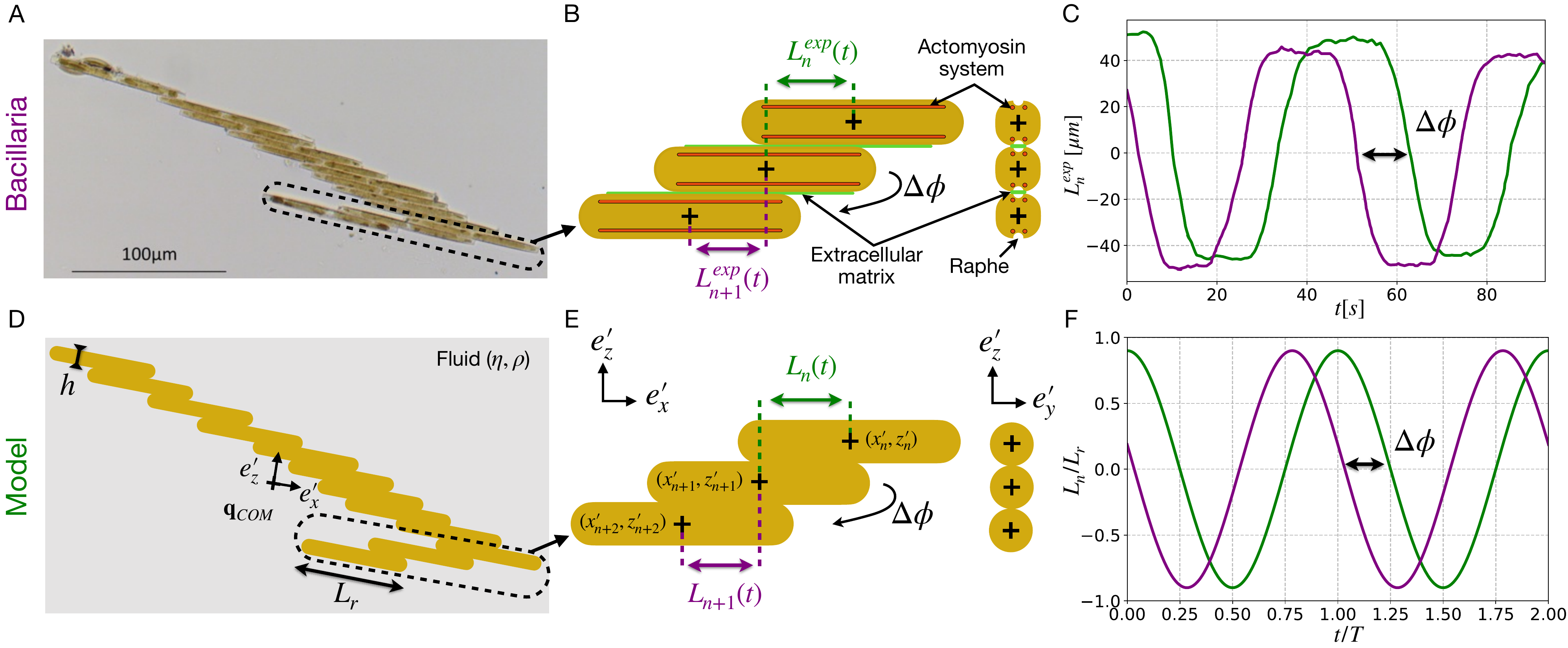}
\caption{Sliding-based actuation inspired by   \textit{Bacillaria}.  
(A) Light microscopy image of a twelve-cell colony of \textit{Bacillaria paxillifer} (adapted from \cite{iNaturalist_photo_11624974}). 
(B) Schematic of the diatom chain architecture. Left: side view. Each cell contains two longitudinal slits, called raphes,  that enable adhesion through secreted strands of an extracellular matrix. These strands are coupled to intracellular actin filaments via myosin motors, producing relative sliding between neighboring cells. Right: schematic of the colony cross section showing the raphes. 
(C) Experimentally measured relative displacement $L_n^{\mathrm{exp}}$ between the geometric centers of adjacent diatoms (green and purple curves), showing the phase shift $\Delta\phi$ between different cell pairs (adapted from \cite{Yamaoka_colonial_2016}). 
(D) Schematic of the bio-inspired model used in numerical simulations, consisting of $N_{\mathrm{rods}}$ rigid rods of length $L_r$ and diameter $h$.
$\mathbf{q}_{\mathrm{COM}}$ denotes the colony center of mass, and $(\mathbf{e}'_x,\mathbf{e}'_y,\mathbf{e}'_z)$ defines the body-fixed reference frame. 
(E) Illustration of the imposed relative sliding between adjacent rods in the model. Left: side view in the ($x',z'$) plane. $L_n(t)$ denotes the tangential displacement, i.e.\ along $\mathbf{e}'_x$, between rods $n$ and $n+1$, with a prescribed phase shift $\Delta\phi$ between neighboring pairs. Right: schematic of the circular cross-section of rods in the  ($x',y'$) plane. 
(F) Temporal evolution of the normalized relative displacement $L_n/L_r$, modeled as a sinusoidal function (Eq.\ \eqref{eq:sliding}) to mimic the periodic sliding observed experimentally in panel C. }

\label{fig:figure1}
\end{figure*}

Some \removed{pennate} diatoms are capable of sliding not only on surfaces such as soils or substrates but also between neighboring cells. A notable example is \textit{Bacillaria} colonies, which form long chains of elongated rod-shaped cells (Fig.\ \ref{fig:figure1}A) \cite{Mueller_kleine_1782}. Individual cells adhere to one another through the extracellular adhesive material and slide relative to their neighbors via the actin–myosin–driven gliding mechanism (Fig. \ref{fig:figure1}B) \cite{Yamaoka_colonial_2016,Schmid_paradox_2007}. \textit{Bacillaria} colonies can consist of large numbers of attached cells, sometimes reaching several dozen \cite{Yamaoka_colonial_2016}. The relative sliding motion observed between adjacent cells is often periodic, with neighboring pairs of rods oscillating with a phase shift (Fig. \ref{fig:figure1}C) \cite{Yamaoka_colonial_2016,harbich2023modeling}. The resulting collective motion of the chains is therefore also periodic, producing traveling waves reminiscent of the planar beating of eukaryotic flagella. In classical flagellar systems, propulsion occurs opposite to the direction of the traveling wave
and the swimming speed exhibits a single optimum, typically corresponding to roughly 1-1.5  oscillation wavelengths along the flagellum \cite{higdon1979hydrodynamic,Spagnolie_optimal_2010,Lauga_eloy_shape_2013,ishimoto2016hydrodynamic,dresdner1981relationships,Brokaw_non-sinusoidal_1965}.

In this work, we draw inspiration from \textit{Bacillaria} colonies and use theory together with three-dimensional numerical modeling to investigate locomotion driven by sliding between elongated cells. We show that sliding-induced swimming generates a rich set of dynamics that, to our knowledge, has not been previously identified. First, we demonstrate that, unlike flagellar beating, the direction of propulsion is governed by the oscillation wavelength along the colony. When the wavelength falls below a threshold value, the colony moves \textit{in the same direction} as the traveling wave, producing a ``backward-swimming" mode that contrasts sharply with undulatory propulsion, where motion typically  occurs opposite to wave propagation.

Second, we find that this backward mode can be up to 3.5 times faster than forward motion, with each sliding cycle displacing the colony by roughly 1.4 body lengths. We show that this regime emerges from strong rotations generated by sliding-induced shear along the chain. Because of shape asymmetry, these rotations translate into rapid backward propulsion. Beyond being faster, the backward-swimming mode is also the most energetically efficient, independent of colony size and provided the cells are sufficiently elongated. As the cell aspect ratio decreases, the system behaves more like a slender flagellum, and the optimal propulsion mode transitions to the forward-swimming regime typical of beating flagella. Finally, the aspect ratio that maximizes energetic efficiency matches experimental values reported for \textit{Bacillaria} colonies,  suggesting that hydrodynamic efficiency may represent an  evolutionary selective pressure in diatom chains.

Together, these findings establish sliding as a powerful and previously underappreciated mechanism for locomotion in multicellular assemblies. Beyond their biological implications, our results uncover a new class of propulsion strategies that could guide the development of next-generation biomimetic microswimmers and swarm robotic systems, in which robust, coordinated motion and self-assembly emerge from simple local interactions  among elementary units \cite{Das_force_2022, Wang_cilia_2022, Brambilla_swarm_2013, Gardi_microrobot_2022, Xie_reconfigurable_2019, ju2025technology}.

\section*{Results}
\subsection*{Bio-inspired sliding system}

Our \textit{Bacillaria}-inspired model consists of a chain of $N_{\text{rods}}$ stacked rigid rods, of length $L_r$ and diameter $h$, so that the total colony length is $L_c = N_{\text{rods}}\times h$ (Fig.\  \ref{fig:figure1}D). 
To mimic the motion of \textit{Bacillaria} colonies, the rods are linked through kinematic constraints that \textit{(i)} maintain a constant normal separation, \textit{(ii)} enforce parallel alignment, and \textit{(iii)} prescribe a tangential sliding displacement.   
This prescribed sliding motion provides a simplified approximation of the experimentally observed behavior (Fig.\ \ref{fig:figure1}C), represented as a sinusoidal relative displacement $L_n(t)$ between the center of mass of adjacent rods $n$ and $n+1$ (Fig.\ \ref{fig:figure1}F)
\begin{equation}
L_{n}(t)  \equiv  x'_{n+1}(t)-x_n'(t) = A \sin(2 \pi f t + \phi_n),
\label{eq:sliding}
\end{equation}
 where $x_n'(t)$ is the tangential position of the center of mass of rod $n$ in the comoving frame of the colony $(\mathbf{e}'_x(t),\mathbf{e}'_y(t),\mathbf{e}'_z(t))$ (Fig.\ \ref{fig:figure1}D-E),  $A=0.84L_r$
 is the sliding amplitude typically observed in experiments \cite{kapinga1992cell,Yamaoka_colonial_2016},  
$f=1/T$ is the sliding frequency, 
and $\phi_n = (n-1) \Delta \phi$ is the phase of the $n$-th pair of rods, with $\Delta \phi$ the phase shift between adjacent pairs. Despite the variability observed at the biological scale, we assume here that $A$ and $\Delta \phi$ are uniform across the colony to simplify the modeling.

   
Given the very low Reynolds number at the colony scale ($Re \ll 1$), inertial effects are negligible compared to viscous forces \cite{Purcell_life_1977,lauga_hydrodynamics_2009}. 
The surrounding quiescent fluid is therefore governed by the Stokes equations, with no-slip boundary conditions imposed on the surfaces of the sliding rods, and gravitational effects neglected. 
To determine the colony motion induced by this sliding activity, we compute the three-dimensional hydrodynamic interactions between rods using the rigid multiblob method \cite{balboa_usabiaga_hydrodynamics_2016}, which has been widely employed to model complex suspensions \cite{Delmotte_rigidmultiblob_2025}, including rod-like particles.  In this method, each rod is discretized with markers distributed along its centerline, with a spacing chosen to match highly accurate predictions of its mobility coefficients  \cite{balboa_usabiaga_hydrodynamics_2016}.  Further details on the governing equations and numerical implementation are provided in Section B of the Supplementary Material and in previous works \cite{Usabiaga_numerical_2022,Delmotte_rigidmultiblob_2025}.

 The relative motion between pairs of rods can be interpreted as a deformation wave traveling backwards in the frame of the colony (see Fig.\ \ref{fig:figure1}E and \ref{fig:figure2}A):
 $L(z'_n,t)   = A \sin\big[ k (z'_n + c t)\big] $, with wave number $k = 2 \pi / \lambda = \Delta \phi /h$, wave speed $c = 2 \pi f / k$, and discrete vertical positions $z'_n = (n-1) h$.
  Using this traveling wave equivalence, the oscillatory dynamics of a sliding chain can be quantified with the number of wavelengths along the colony $N_\lambda = \frac{L_c}{ \lambda} = \frac{(N_{\rm rods}-1)\Delta \phi}{2\pi} = \frac{\Delta \Phi}{2\pi}$, which can also be interpreted as the total sliding phase shift along the colony $\Delta \Phi$ normalized by a full oscillation $2\pi$. 
  In the following, 
  lengthscales are non-dimensionalized with the colony size $L_c$, time scales with  the sliding period $T$, and forces with  $\mu L^2_c/T$, where $\mu$ is the fluid viscosity.

\subsection*{Sliding-induced swimming is  bidirectional}

\begin{figure*}[]
\centering
\includegraphics[width=0.75\linewidth]{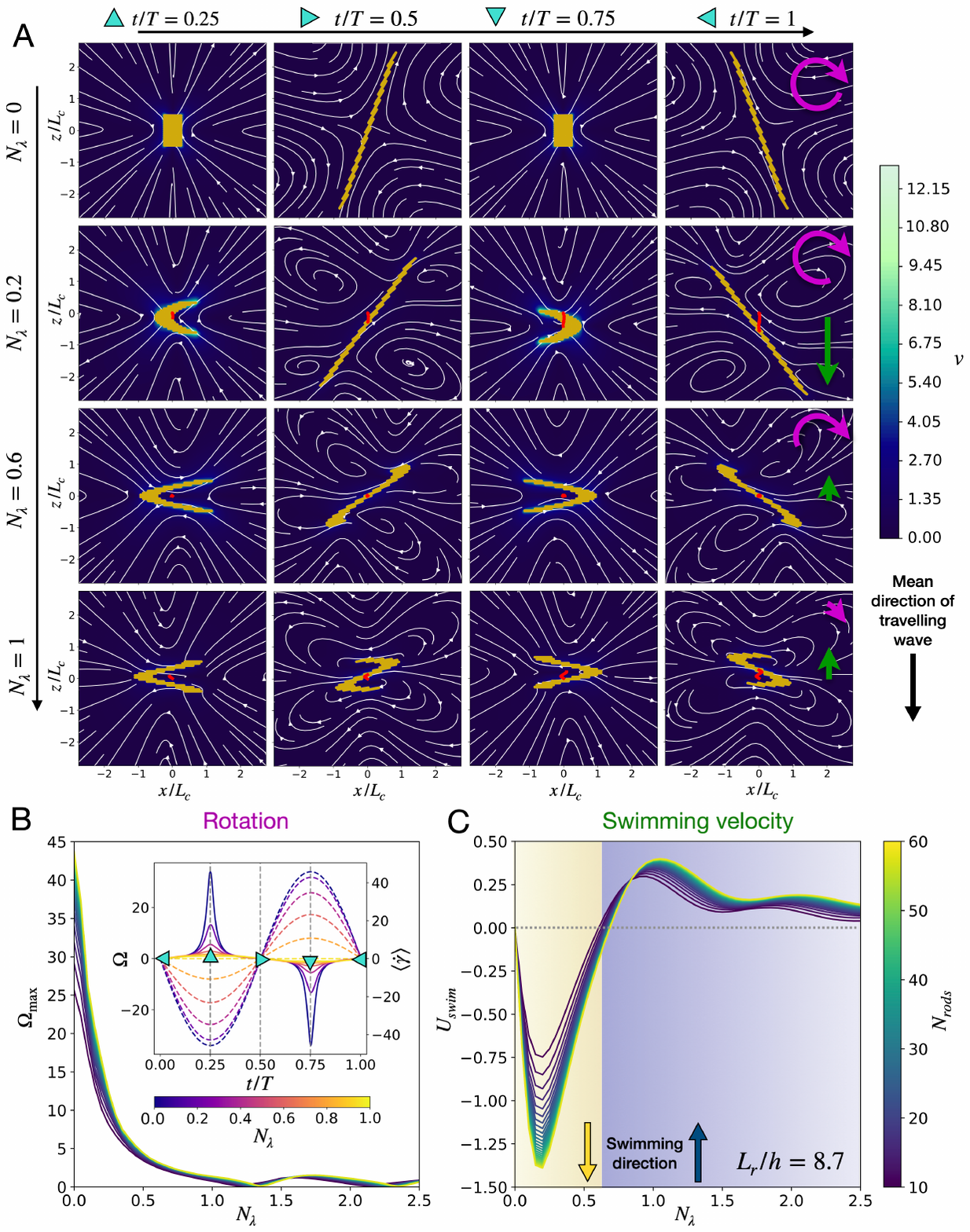}
\caption{Swimming dynamics and motion reversal. (A) Simulated three-dimensional flow field in the mid-plane of a chain  with  $N_{\rm rods} = 16$ rods of aspect ratio $L_r/h = 8.7 $ (see also Supplementary Movies). Rows correspond to different numbers of oscillations along the colony, $N_{\lambda}$, and columns show successive instants within one oscillation period. Red line:  trajectory of the colony  center of mass (COM). Background color:  fluid velocity field $v$. White lines: flow streamlines. Purple arrow: typical magnitude of the rotation rate. Green arrow: direction  and typical magnitude of the swimming velocity.  (B) Maximum rotation rate $\Omega_{\max}$ attained within one sliding period as a function of the number of oscillations  $N_{\lambda}$ for a fixed aspect ratio  $L_{r}/h = 8.7$  and colony sizes in the range   $N_{\rm rods} = 10 - 60$, as indicated by the colorbar.   Inset: Instantaneous rotation rate $\Omega(t)$ (solid lines) and  mean shear along the colony $\langle \dot{\gamma}\rangle(t)$ (dashed lines)  over one  sliding period for $N_{\rm rods} = 16$ and $L_r/h = 8.7$. Color indicates $N_{\lambda} \in [0,1]$. The four markers along the darkest curve correspond to the four snapshots shown in panel A for   $N_{\lambda}=0$.  (C) Swimming velocity averaged over one period, $U_{\text{swim}}$, as a function of  $N_{\lambda}$ for  an aspect ratio  $L_{r}/h = 8.7$ and colony sizes in the range   $N_{\rm rods} = 10 - 60$. Color denotes colony size $N_{\rm rods}$. Yellow and purple shaded regions indicate backward- and forward-swimming modes, respectively, as indicated by the thick arrows.}
\label{fig:figure2}
\end{figure*}

The swimming dynamics of the colony, obtained from our three-dimensional numerical simulations, as a function of $N_\lambda$ is investigated and illustrated in Fig.\ \ref{fig:figure2} (see also SI  Movies). Since the sliding motion is planar and periodic, the dynamics is measured in the frame of the time-averaged orientation vectors $\mathbf{e}_z = \frac{1}{T} \int_0^T \mathbf{e}'_z(t)\, \dd t$, and $\mathbf{e}_x = \frac{1}{T} \int_0^T \mathbf{e}'_x(t)\, \dd t$.
In this frame, we find that there is not net motion along $\mathbf{e}_x$ and the mean direction of the traveling wave is $-\mathbf{e}_z$.
Figure \ref{fig:figure2}A presents the center-of-mass trajectories, colony conformations, and surrounding flow fields over a swimming cycle for four values of $N_\lambda$.  When $N_\lambda = 0$, there is no phase shift between rod pairs, and all rods slide in synchrony. Consequently, the colony undergoes pronounced rotation but no net translation, as the motion is fully symmetric and time-reversible. Increasing $N_{\lambda}$ to 0.2 breaks this symmetry: although substantial rotation persists, the colony now translates by roughly one body length per cycle in the direction of wave propagation, that is, backwards. At $N_{\lambda}=0.6$, rotational motion is markedly reduced, the mean swimming speed becomes very small, and the net displacement reverses direction with a slight forward motion, opposite to the wave direction. When $N_{\lambda}=1$, rotation diminishes further, and the colony translates against the traveling wave, consistent with classical undulatory propulsion.

The dependence of the rotational and translational motions on $N_{\lambda}$ is summarized in Figs. \ref{fig:figure2}B and \ref{fig:figure2}C respectively. 
The peak angular velocity within a sliding cycle, $\Omega_{\max}$, is highest at $N_{\lambda} = 0$ and decreases rapidly as $N_{\lambda}$ increases, with only minor oscillations appearing for $N_\lambda \gtrsim 1.3$, independent of the colony size  $N_{\rm rods}$ (Fig.\ \ref{fig:figure2}B). 
To relate the colony’s rotation to the imposed sliding motion, we compute the mean shear along the chain in the comoving frame, $\langle \dot{\gamma}\rangle(t)$, defined as the difference in sliding velocity $u_{\text{slide}}(t)$ between the two end  rods divided by their distance  (see Supplementary material Sec.\ C.1-C.2 for derivations of the mean shear along the colony). 
The inset of Fig.\ \ref{fig:figure2}B plots the rotation rate $\Omega(t)$ (solid line) and the mean shear along the colony   $\langle \dot{\gamma}\rangle(t)$ (dashed line) over one sliding period  for various  $N_{\lambda}$ and $N_{\rm rods}=16$. 
At $N_{\lambda} = 0$, the distal rods slide in opposite directions with large amplitudes, producing maximal shear and thus 
the strongest rotation. At $N_{\lambda} = 1$, or any other positive integer, the colony spans a full wavelength, yielding equal sliding velocities at both ends, vanishing mean shear, and thus minimal rotation. For intermediate $N_{\lambda}$, the velocity difference between distal rods decreases with $N_\lambda$, leading to progressively lower shear and weaker rotation.

In contrast to rotation, the normalized mean swimming speed  $U_{\text{swim}}$,  directed along the $z-$axis since the $x-$component averages to zero, varies strongly and non-monotonically with $N_{\lambda}$, following the same trend across all colony sizes  (see Fig.\ \ref{fig:figure2}C). Because the motion is reciprocal at $N_{\lambda}=0$, the swimming speed is initially zero, but it quickly reaches a first negative maximum (backward propulsion) near $N_\lambda \approx 0.2$, approaching $U_{\text{swim}}\approx 1.4$ colony lengths per sliding period for $N_{\rm rods} = 50$. A second, much smaller maximum, roughly 3.5-fold weaker, occurs at $N_\lambda\approx 1.1$ and corresponds to forward swimming,  opposite to the traveling wave. For larger $N_\lambda$, progressively weaker forward peaks appear periodically. The transition between backward and forward propulsion occurs at $N_{\lambda}\approx 0.6$.


  
\subsection*{Backward swimming is faster due to self-induced shear  and rotation-translation couplings}
\begin{figure*}[t]
\centering
\includegraphics[width=0.9\linewidth]{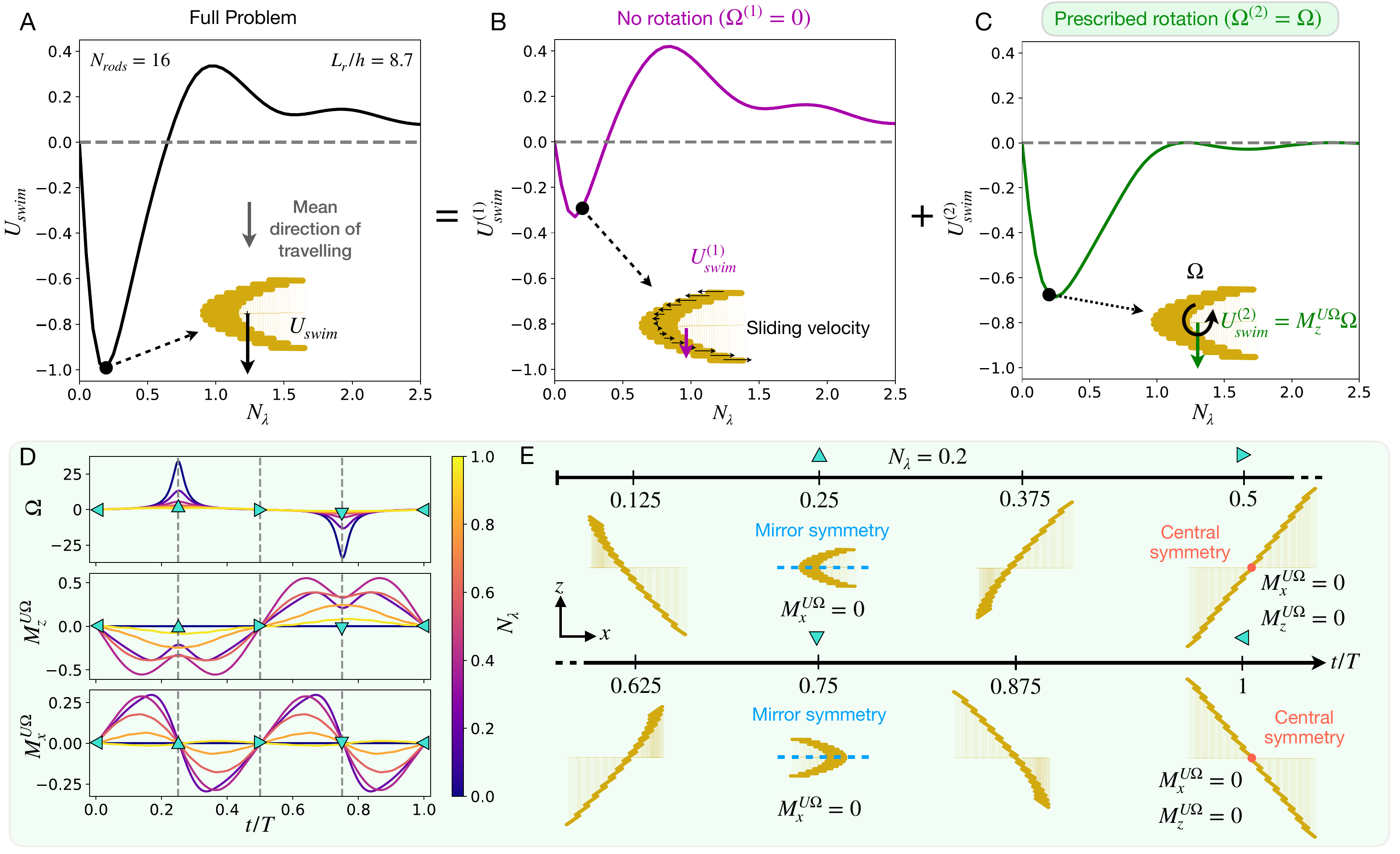}
\caption{ Mechanism of backward swimming. 
(A) Swimming velocity averaged over one period, $U_{\mathrm{swim}}$, as a function of the number of oscillations along the colony, $N_{\lambda}$, for $N_{\mathrm{rods}} = 16$ and aspect ratio $L_r/h = 8.7$. The inset shows a typical colony deformation for the fastest backward-swimming mode $N_{\lambda}=0.2$ indicated by the black disk.
(B, C) Decomposition of the full swimming problem in panel A into two subproblems: (B) motion with rotation suppressed $(\Omega^{(1)} = 0)$ and (C) motion with prescribed rotation $(\Omega^{(2)} = \Omega)$ on the   colony treated as a rigid body. 
(D) Time evolution evolution of the rotation-tranlsation coupling coefficients $M^{U\Omega}_{z}(t)$ and $M^{U\Omega}_{x}(t)$ in subproblem 2 together with the angular velocity $\Omega(t)$  over one sliding period for  $N_{\mathrm{rods}} = 16$ and rod aspect ratio $L_r/h = 8.7$. 
Curves are colored according to the wavelength parameter $N_{\lambda} \in [0,1]$. The markers indicate the four instants corresponding to the snapshots shown in Fig.~2A.
(E) Chronophotographs of the colony in the ($x-z$)  frame for $N_{\lambda}=0.2$, $N_{\mathrm{rods}} = 16$ and aspect ratio $L_r/h = 8.7$, illustrating the evolution of the configuration over one period. Mirror  and central symmetries are highlighted. 
}
\label{fig:figure3}
\end{figure*}

To elucidate the origin of the bidirectional swimming behavior and the emergence of a fast backward-propulsion mode, we exploit the linearity of Stokes flow to decompose the swimming problem, at each instant within a sliding period, into two subproblems (Fig. \ref{fig:figure3}A–C). 
For a given instantaneous geometric configuration, the first subproblem corresponds to the swimming problem with rotation suppressed ($\Omega^{(1)} = 0$, $L_n(t)\ne0$). In the second subproblem, the colony is treated as a single rigid body of identical shape, i.e.\ without sliding motion, and is prescribed to rotate exactly as in the full swimming problem ($\Omega^{(2)} = \Omega(t)$,  $L_n(t)=0$). Full equations and details of the linear superposition  are provided in Section D of the Supplementary Material.  


 
When the instantaneous angular velocity is canceled (subproblem 1), sliding motion predominantly induces forward propulsion ($U^{(1)}_{\rm swim}>0$ for $N_{\lambda}<0.4$), with a local maximum near $N_\lambda \approx 1$, consistent with classical flagellar locomotion (see Fig.~\ref{fig:figure3}B)~\cite{higdon1979hydrodynamic}. In the comoving frame of the colony, propulsion is directed exclusively in the forward direction (see SI Fig.~3B), i.e.\ along $\mathbf{e}'_{z}$, and its magnitude is accurately estimated using Resistive Force Theory  (RFT)~\cite{hancock1953self,gray1955propulsion,cox1970motion}, in which the colony centerline is modeled as an undulating filament subject to anisotropic drag (see Supplementary Material Sec.~E for details). The backward motion observed for $N_{\lambda}<0.4$ arises because the colony is strongly rotated at small $N_{\lambda}$ (see Fig.~\ref{fig:figure2}B), causing the deformation wave, when expressed in the colony frame, to yield net backward propulsion in the mean swimming frame (see SI Fig.\ 2).
 
In the second subproblem, locomotion arises solely from the imposed rotation on the rigid colony (Fig.~\ref{fig:figure3}C).
Because of the colony's geometric asymmetry, imposed rotation couples to translation, producing a net backward motion whose speed approaches twice the maximum forward velocity attainable without rotation. To quantify this coupling, we use our numerical method to evaluate the coupling coefficients, (${M}^{U\Omega}_x(t)$, ${M}^{U\Omega}_z(t)$), which  give the instantaneous  translational velocity of subproblem 2  ($U^{(2)}_x(t)$, $U^{(2)}_z(t)$) in response to  the prescribed rotation on the rigid colony with a given shape at time $t$:  $U^{(2)}_x(t) = {M}^{U\Omega}_x(t)\Omega(t)$  and $U^{(2)}_z(t) = {M}^{U\Omega}_x(t)\Omega(t)$. Because of the linearity of Stokes flow, these two coefficients only depend on the instantaneous shape of the colony  (see Supplementary Material Sec. D.3 for details on the method to compute ${M}^{U\Omega}$). 
Fig.\ \ref{fig:figure3}D shows the evolution of $M^{U\Omega}_{z}(t)$ and $M^{U\Omega}_{x}(t)$  together with the angular velocity $\Omega(t)$  over one sliding period for  $N_\lambda \in [0,1]$. 
We first observe that $M^{U\Omega}_{z}(t)$ is consistently opposite in sign to $\Omega(t)$ and has the same period $T$. This explains why the rotation-induced velocity averaged over one period is directed backward, $U^{(2)}_{\rm swim} = 1/T\int_0^TM^{U\Omega}_{z}(t)\Omega(t)\dd t < 0 $. 
In contrast, $M^{U\Omega}_{x}(t)$ has a period half that of $\Omega(t)$. This period doubling results in a vanishing net velocity along the $x$-direction over one sliding period, $\int_0^TM^{U\Omega}_{x}(t)\Omega(t)\dd t = 0$.

To elucidate the distinct behaviors of the two coupling coefficients, we examine the symmetries of the colony shape over one period in Fig.\ \ref{fig:figure3}E. Every quarter period, the colony alternates between a mirror-symmetric shape about the $x$-axis, at $t=0.25T$ and $0.75T$, 
and a centrally symmetric shape, at $t=0.5T$ and $T$. 
By virtue of the reversibility of Stokes flow \cite{happel2012low}, rotation of a centrally symmetric shape about the $y$-axis cannot induce translation ($M^{U\Omega}_{x}=M^{U\Omega}_{z}=0$). For a shape possessing mirror symmetry about the $x$-axis, however, rotation gives rise to translation solely along the $z$-direction ($M^{U\Omega}_{x}=0, M^{U\Omega}_{z}\neq 0$). 
This symmetry argument explains why $M^{U\Omega}_{x}$ changes sign and cancels every quarter period, twice as often as $M^{U\Omega}_{z}$.
As shown in Fig.\ \ref{fig:figure2}A, this alternation between mirror-symmetric and centrally symmetric shapes holds for any wavelength $N_{\lambda}$.

Finally, we observe in Fig.\ \ref{fig:figure3}D that rotation weakens with increasing $N_{\lambda}$, since the mean shear decreases (see inset of Fig.\ \ref{fig:figure2}B), whereas $M^{U\Omega}{z}(t)$ exhibits a non-monotonic dependence on $N_{\lambda}$ and reaches a maximum near $N_{\lambda}\approx 0.4$. These competing trends explain the non-monotonic dependence of $U^{(2)}_{\rm swim}$ on $N_{\lambda}$ and the existence of a maximum velocity at $N_{\lambda}\approx 0.2$, where the product $M^{U\Omega}_{z}(t)\Omega(t)$ is maximized over one sliding period.

Overall, these results demonstrate that the reversal in swimming direction can be interpreted as a competition between two distinct propulsion mechanisms arising from inter-rod sliding: a flagellar-like, predominantly forward, propulsion driven by a deformation wave propagating along the centerline (subproblem 1), and a strong rotation-translation coupling due to shear and colony-scale asymmetry, which drives exclusively backward motion (subproblem 2). Their combination gives rise to the fast backward-swimming mode observed at $N_{\lambda} \approx 0.2$.

\subsection*{Backward motion is optimally efficient}

\begin{figure*}[t]
\centering
\includegraphics[width=0.9\linewidth]{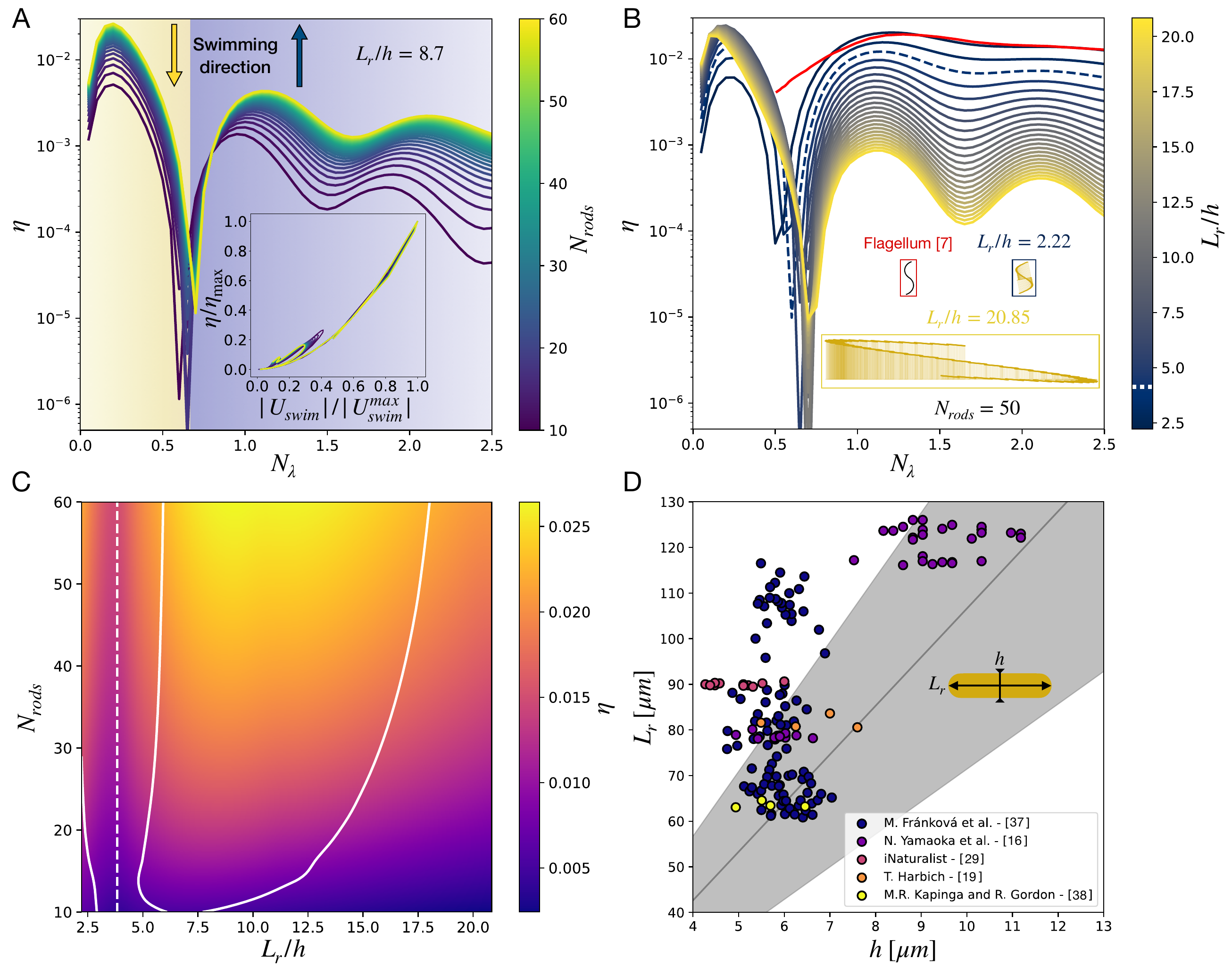}
\caption{Swimming efficiency.
(A) Swimming efficiency, $\eta$, as a function of the number of oscillations, $N_{\lambda}$, for a fixed rod aspect ratio $L_r/h = 8.7$. Color denotes the colony size $N_{\rm rods} = 10 - 60$. Inset: efficiency $\eta$ \textit{vs.} swimming speed $U_{\rm swim}$ normalized by their respective maxima for $N_{\rm rods} = 10 - 60$.  
(B) Swimming efficiency, $\eta$, as a function of $N_{\lambda}$ for a fixed colony size $N_{\mathrm{rods}} = 50$. Color denotes the rod aspect ratio $L_r/h = 2.2-20.85$. The red curve shows the data for a finite slender flagellum (see red inset)  reported by Higdon \cite{higdon1979hydrodynamic} (these data have been scaled by a constant prefactor of 2.1 to match the absolute value of the computed efficiency). The dashed dark blue curve at $L_r/h \approx 4$ marks the transition from an optimal forward-swimming mode $(N_{\lambda} \approx 1.2)$ to an optimal backward-swimming mode $(N_{\lambda} \approx 0.2)$. Black and yellow insets show snapshots of the colony for $N_{\lambda} = 1$  at two aspect ratios, $L_r/h = 2.2$ and 20.85.
(C) Maximal swimming efficiency as a function of colony size $N_{\mathrm{rods}}$ and rod aspect ratio $L_r/h$. White solid contours delineate regions where the efficiency is within 85$\%$ of the maximum for a given colony size. The white dashed line at $L_r/h \approx 4$ indicates the minimum efficiency for each colony size and corresponds to the transition between optimal forward- and backward-swimming modes.
(D) Diatom cell length as a function of cell height, compiled from experimental measurements in the literature \cite{Frankova_halophilous_2007,Yamaoka_colonial_2016,iNaturalist_photo_11624974,harbich2023modeling,Kapinga_cell_1992}. The dark gray solid curve indicates the mean aspect ratio predicted to maximize swimming efficiency ($L_r/h=10.66)$, and the light gray band denotes its standard deviation  ($\sigma = 3.52$).}
\label{fig:figure4}
\end{figure*}

To assess whether the fast swimming modes are also energetically favorable, we compute the colony's swimming efficiency $\eta$, commonly referred to as the Froude efficiency \cite{guastoplanktonic2012}. For sliding-driven motion, the swimming efficiency is defined as the ratio of the period-averaged power required to tow the colony at its mean swimming speed to the total power expended over one sliding cycle to generate the motion:

\begin{equation}
\eta=\frac{P_{\text {pull }}}{P_{\text {swim }}}
\end{equation}
with
\begin{equation}
P_{\text {pull }}= \frac{1}{T} \int_0^T \boldsymbol{R}^{F U}(t) \dd  t \cdot {{\mathbf{U}}_{\rm swim} \cdot {\mathbf{U}}_{\rm swim}}
\end{equation}
and
\begin{equation}
P_{\text {swim }}=\frac{1}{T} \int_0^T \left(\begin{array}{c}
\boldsymbol{f}_{\text{slide}}(t)\\
\boldsymbol{\tau}_{\text{slide}}(t)\\
\end{array}\right) \cdot \left(\begin{array}{c}
\boldsymbol{u}(t)\\
\boldsymbol{\omega}(t)\\
\end{array}\right) \dd t
\end{equation}
where $\boldsymbol{R}^{F U}(t)$ is the resistance matrix  that provides the drag for a given colony  shape at  instant $t$, ($\boldsymbol{f}_{\rm slide}(t)$, $\boldsymbol{\tau}_{\text{slide}}(t)$) are the  forces and torques applied to the rods to generate the desired sliding motion, and ($\boldsymbol{u}(t)$, $\boldsymbol{\omega}(t)$) are their translational and rotational velocities respectively.

We find that swimming efficiency mirrors the trend observed for mean swimming speed across all colony sizes (Fig. \ref{fig:figure4}A): the fastest mode, occurring at  $N_\lambda=0.2$, is also the most efficient one; efficiency drops to zero at the backward–forward transition ($N_{\lambda}\approx0.6$) and exhibits secondary maxima, corresponding to forward propulsion modes, at identical values of $N_\lambda$.  
Interestingly, the overlapping curves in the inset of Fig. \ref{fig:figure4}A, which shows  $\eta$ \textit{vs.} $U_{\rm swim}$ normalized by their respective maxima, indicate that these quantities are not only correlated but also approach an asymptotic behavior as the number of rods  $N_{\rm rods}$ increases, suggesting a universal dependence in the large-colony limit. 
While the number of cells varies widely among diatom colonies \cite{Yamaoka_colonial_2016}, our results indicate that colony size has little influence on the swimming efficiency of our bio-inspired chains. 
In contrast, the cell aspect ratio, $L_r/h$, plays a key role: it controls the amplitude of intercellular sliding, the resulting shear strength, and the overall shape anisotropy of the colony.

To examine the effect of cell aspect ratio on swimming efficiency, we fix the number of rods $\left(N_{\text {rods }}=50\right)$ and vary  $L_r / h \in$ [2.2, 20.8]. 
As shown in Fig.\ \ref{fig:figure4}B, the qualitative evolution of swimming efficiency varies greatly with cell aspect ratio: the range of $N_{\lambda}$ producing backward motion narrows for smaller $L_r/h$,  and  more importantly, the optimal wavelength shifts from $N_\lambda \approx 0.2$ (backward swimming) for $L_r / h>4$, to $N_\lambda \approx 1.2$ (forward swimming) for $L_r  / h<4$.
The optimal value $N_\lambda \approx 1.2$ closely matches observations in flagellated microorganisms, where the most efficient stroke spans approximately one wavelength \cite{dresdner1981relationships,Spagnolie_optimal_2010,tam2011optimal,Lauga_eloy_shape_2013,ishimoto2016hydrodynamic}, and qualitatively agrees with Higdon’s prediction for a finite slender flagellum undergoing a sinusoidal traveling wave \cite{higdon1979hydrodynamic}  (see red line in Fig.\ \ref{fig:figure4}B). This similarity to flagellar dynamics at low aspect ratios is expected, as the colony resembles more as a slender filament when $L_r/h$  drecreases (see inset of Fig.\ \ref{fig:figure4}B). 

To summarize the influence of colony geometry, Fig.\ \ref{fig:figure4}C shows the maximum swimming efficiency $\eta_{\max}(N_{\rm rods},L_r/h) = \max_{N_{\lambda}} \eta(N_{\lambda},N_{\rm rods},L_r/h)$ as a function of rod aspect ratio and colony size. 
The peak efficiency reaches approximately $2.5\%$, which is consistent with the values expected for microorganisms \cite{Purcell_life_1977,Chattopadhyay_swimming_2006,Omori_swimming_2020}. 
The low efficiency observed near $L_r/h\approx 4$ corresponds to the transition between forward and backward optimal modes  at  $N_{\lambda}\approx1.2$  and $N_{\lambda}\approx 0.2$, respectively (see  Sec.\ F in the Supplementary Material).
The white contour in Fig.\ \ref{fig:figure4}C highlights the region where efficiency is within $15\%$ of the local maximum for a given colony size.
By analyzing the distribution of aspect ratios within this high-efficiency region, we find a mean optimal aspect ratio of $L_r/h=10.66$ with a standard deviation of $\sigma = 3.52$ (see full distribution in  Supplementary Material Sec.\ G).  
Figure \ref{fig:figure4}D presents measurements of cell lengths and heights extracted from various experimental datasets and published images (see Supplementary Material Sec.\ H for extraction method). Most experimental values lie within the range of optimal aspect ratios predicted by our model, supporting the relevance of these geometric constraints for efficient swimming.

\section*{Discussion}
Our study shows that sliding-driven actuation in diatom-inspired chains yields a previously unrecognized swimming mechanism in which both the direction and efficiency of propulsion depend on the number of oscillations along the colony. In contrast to classical flagellar propulsion, which relies on  bending waves along the centerline, sliding  introduces an additional degree of freedom which modulates the shear induced on the colony, giving rise to two competing propulsion modes: forward motion driven by traveling-wave-like centerline deformations, and backward motion arising from shear-induced rotation created by sliding. Notably, the backward-swimming mode is both faster and more efficient than the traditional forward mode.

Our parametric analysis of chain geometry identified an optimal aspect ratio between 7 and 14, consistent with values reported for \textit{Bacillaria} colonies. While this correspondence is intriguing, it does not necessarily imply that diatom morphology is hydrodynamically optimized toward this ratio. Microorganisms typically evolve under multi-objective, Pareto-like constraints rather than through single-criterion optimization \cite{Schuech_motile_2019}. Moreover, elongated rod-shaped cells may provide ecological benefits—such as improved vertical migration, enhanced stability in turbulent flows \cite{Lovecchio_chain_2019}, and more effective pairing or collective sinking behaviors observed in diatoms \cite{Font-Munoz_collective_2019}. Thus, although hydrodynamic efficiency may play a role in shaping these morphologies, it is likely only one of several selective pressures acting in concert.

Beyond probing the swimming dynamics of diatom-like chains, our model offers a framework for investigating sliding mechanisms in real colonies in conjunction with experiments. Experimentally measured sliding motions can be imposed in the model, allowing direct comparison between the simulated and observed colony-level motions. The sliding forces required to reproduce these motions in silico would provide noninvasive estimates of the intercellular forces and of the cell–substrate interactions involved. Such quantitative estimates could in turn clarify the underlying biophysical processes.

In addition to cell–cell sliding, \textit{Bacillaria} colonies sometimes anchor to a substrate by one of their end cells. In this configuration, the cells lie parallel to the surface and form a stack in the normal direction while performing periodic sliding motions. This behavior currently lacks a mechanistic explanation, and our model could help clarify its functional role by quantifying the flows and pumping rates produced by these tethered sliding patterns.


Beyond their biological significance, our results suggest new design principles for bio-inspired microswimmers and collective robotic systems, in which  interactions between elementary units with few degrees of freedom and simple sensory feedback can give rise to self-assembly and coordinated motion at the colony or swarm scale \cite{Das_force_2022, Wang_cilia_2022, Brambilla_swarm_2013, Gardi_microrobot_2022, Xie_reconfigurable_2019, ju2025technology}. By leveraging optimal control strategies and reinforcement learning approaches \cite{moreau2023controllability,xiong2025chemotactic, liu2025reinforcement}, the geometry of the chain elements  and their sliding motion could be tuned to achieve energy-efficient propulsion or to autonomously perform specific tasks, such as directed navigation, environmental sensing, fluid mixing, pumping, or targeted cargo transport.\\


\paragraph*{Acknowledgements.} 
The authors thank Gabriel Amselem and Thomas Harbich for many insightful discussions.
 BD acknowledges support from the French National Research Agency (ANR), under awards ANR-20-CE30-0006 and ANR-25-CE30-4710.

\bibliography{Ref_article}

\end{document}